\documentclass{aa}
\usepackage{epsfig}
\usepackage{float,times}
\usepackage{rotating}
\usepackage{graphics}

\begin{document}

\thesaurus{02 (11.17.1; 12.12.1; 03.13.6)} 

\title{Detecting the non-Gaussian Spectrum of QSO's Ly$\alpha$
Absorption Line Distribution}

\author{Jes\'{u}s Pando \inst{1,}\inst{2} \and Li-Zhi Fang \inst{2}}
\offprints{J. Pando}

\institute{UMR 7550 CNRS, Observatoire de Strasbourg, 
67000 Strasbourg, France
\\ pando@wirtz.u-strasbg.fr
\and Department of Physics, University of Arizona, 
 Tucson, AZ  85721, USA \\
fanglz@physics.arizona.edu}

\date{Received date / Accepted date} 

\maketitle

\markboth{J. Pando and L.-Z. Fang: Non-Gaussian Spectrum of QSO's 
Ly$\alpha$ Lines}{}

\begin{abstract}

We present an analysis of the non-Gaussianity in the distribution of
Ly$\alpha$ forest lines in the QSO absorption spectra. 
Statistical tests performed on this data indicate that there may be 
large scale structure even though the power spectrum of the Ly$\alpha$
line distribution on large scales is found to be flat. It is apparent 
that higher (than two) order statistics are crucial in 
quantifying the clustering behavior of Ly$\alpha$ clouds.

A method of detecting the spectrum of cumulants of the discrete wavelet 
transform (DWT) coefficients is developed. Because the basis of the DWT is 
compactly supported, the DWT is not subject to the central limit theorem.
Cumulants of any order can be quickly computed and serve as a way of detecting
non-Gaussianity. 

Using this technique on three independent data sets of Ly$\alpha$
forests, we find that the distribution of Ly$\alpha$ forests does show
non-Gaussian behavior on scales from 5 to 10 h$^{-1}$ Mpc with confidence 
level larger than 95\%. Two data sets available on large scales are found to 
be non-Gaussian on even larger scales. These techniques are effective in 
discriminating among models of the Ly$\alpha$ forest formation, which are 
degenerate at second and lower order statistics.

\keywords{cosmology: large-scale structure of
Universe  -- quasars: absorption lines}

\end{abstract}

\section{Introduction}

Ly$\alpha$ absorption line forests in QSO spectra come from intervening
absorbers, or clouds, with neutral hydrogen column densities ranging from
about $10^{12}$ to $10^{17}$ cm$^{-2}$ at high red-shifts. Since
the size of the Ly$\alpha$ clouds at high red-shift is as large as
100 - 200 h$^{-1}$ Kpc, and their velocity dispersion is as low as
$\sim$ 100 km s$^{-1}$ (Bechtold et al. 1994, Dinshaw et al. 1995,
Fang et al. 1996), it is generally believed that the Ly$\alpha$ forests are
due to the absorption of pre-collapsed clouds in the density field of the
universe. Ly$\alpha$ clouds are probably fair tracers of the cosmic density 
field on large scales and therefore, the clustering behavior of
the Ly$\alpha$ clouds should be useful for testing models of structure 
formation of the universe. 

More importantly, unlike other high redshift objects, Ly$\alpha$ forests
do not show significant power in their two-point correlation functions. 
Aside from very small scales $\Delta v \leq 300$ km/s,
 all results drawn from the two-point correlation function of the
Ly$\alpha$ absorption lines have failed to detect clustering
(Webb, 1987, Weymann 1993, Hu et al. 1995, Cristiani et al. 1997).
The  power spectrum of the 1-D spatial distribution of the Ly$\alpha$ 
absorbers is found to be flat on scales in velocity space of $\sim$ 600 to 
30,000 km s$^{-1}$ (Pando \& Fang 1998). This result  indicates that 
the distribution of the Ly$\alpha$ clouds may still be in the linear or 
quasilinear evolutionary stages on scales larger than a few h$^{-1}$ Mpc. 
Indeed, it is found that simulations of popular models using the linear
or log-normal approximation fit well with the second order statistical 
properties of Ly$\alpha$ forests (Bi, Ge \& Fang 1995, Bi \& Davidson 1997). 
Therefore, the Ly$\alpha$ clouds may contain information of cosmic 
clustering in the linear or quasilinear evolutionary stages.  

It is known that even though the evolution of the power spectrum during the 
quasi-linear regime does not significantly
differ from the linear regime, the density perturbations on different
scales will no longer evolve mutually independently because of the power
transfer of perturbations via mode coupling. For popular models, like the cold
dark matter model, the mode-mode coupling of the quasi-linear evolution leads
to a power transfer from large scales to small ones (Suto \& Sasaki
1991). Numerical studies show that the power transfer is already 
significant on scales of about 50 h$^{-1}$ Mpc at redshift $\sim 2$ 
(Jing et al. 1995). Thus, there should exist non-Gaussianity on scales of a few 
10 h$^{-1}$ Mpc which is the  ``remnant'' of the mode-mode coupling 
of the quasi-linear evolution.

This theory is supported by works based on methods other than
the two-point correlation function.  For instance, the distribution of 
nearest neighbor Ly$\alpha$ line intervals is found to be definitely 
different from a Poisson process (Duncan, Ostriker, \& Bajtlik 1989; Liu 
and Jones 1990). A study using the Kolmogorov-Smirnoff
(K-S) statistic, finds that Ly$\alpha$ absorbers show a deviation from a
uniform random distribution at the $\sim 3\sigma$ significance level
(Fang, 1991). Some observations also indicate the existence of 
$\sim 10$ Mpc void (Dobrzycki \& Bechtold 1991), and deviation from uniform 
distribution on larger scales (Crotts 1987, 1989.) However, this 
individual structure cannot be used for a statistical analysis. 
Using a method based on cluster identification, many structures have been 
systematically identified and formed into an ensemble. It is found that 
the abundance of the 
identified ``clusters" with respect to the 
richness are significantly different from a  Gaussian process (Pando \& 
Fang 1996, hereafter PF). Recently, we have also found that the Ly$\alpha$ 
forest line distribution shows significant scale-scale correlations. As a 
consequence
models which predict a Gaussian process for the evolution of the
Ly$\alpha$ clouds are ruled out, and the halos hosting the clouds must have
gone through a ``history'' dependent merging process during their formation
(Pando et al. 1998.) 

 In this paper, we will continue to 
develop the description of the non-Gaussianity of the Ly$\alpha$ line 
distribution. The emphasis of this paper will be to detect the non-Gaussian 
spectrum, and to show its ability to discriminate among  models of Ly$\alpha$ 
cloud formation which are degenerate at second order.

In \S 2, we will describe the observed and simulated samples of Ly$\alpha$ 
forests, and the problems related to their large scale structure detection. 
In \S 3, the DWT technique of non-Gaussian spectrum detection will be
discussed. The results of this analysis for real and simulated samples are
discussed in \S 4. We will show that the distributions of Ly$\alpha$ forest
lines are significantly different from Gaussian distributions. Additionally,
we show that the non-Gaussian spectrum is a powerful tool for distinguishing
between models.

\section{Ly$\alpha$ samples and problems}

In PF, we looked at two popular data sets of Ly$\alpha$ forests. The first  
compiled by Lu, Wolfe and Turnshek (1991, hereafter LWT) contains $\sim$ 950 
lines from the spectra of 38 QSO that exhibit neither broad absorption 
lines nor metal line systems. The second set is from Bechtold (1994,
hereafter JB), which contains  $\sim$ 2800 lines from 78 QSO's spectra, in
which 34 high red-shift QSOs were observed at moderate resolution. In this
paper, we augment those data sets with two observations using the Keck
telescope: 1) Hu et al. (1995, hereafter HKCSR) observed 4 QSO's with a total
of 1056 lines and column density in the range $N_{H \, I} \ge 2 \times 10^{12}$
cm$^{-2}$ at extremely high S/N; 2) Kirkman and Tytler (1997, hereafter KT)
obtained the highest quality spectra published to date from QSO HS 1946+7658
with 466 Ly$\alpha$ forest lines. A typical sample is shown in Figure 1, 
which is a 1-D histogram of the spatial distribution of Ly$\alpha$ absorption 
lines of QSO-0142.

It is well known that the number density of the Ly$\alpha$ absorption lines 
increases with red-shift. The number density of lines with rest equivalent
width $W$ greater than a threshold $W_{th}$ can approximately be described
as
\begin{equation}
\frac{dN}{dz}=\left(\frac{dN}{dz}\right)_0(1+z)^{\gamma} \ ,
\end{equation}
where $(dN/dz)_{0}$ is the number density extrapolated to zero red-shift,
and $\gamma$ the index of evolution. LWT finds that $(dN/dz)_0 \simeq 3$ and 
$\gamma = 2.75 \pm 0.29$ for lines with $W\ge W_{th}=0.36$\AA. KT finds 
$\gamma = 2.6$ while JB finds that $\gamma = 1.89 \pm 0.28$ 
for $W_{th} \ge 0.32$\AA$\;$ and $\gamma=1.32\pm0.24$ for 
$W_{th} \ge 0.16$\AA. 

Like other Ly$\alpha$ forest data, these data sets have failed to reveal 
structures in their distribution on scales $\Delta v \geq 300$ km/s 
when subjected to a two point correlation analysis (Hu et al. 
1995.) Using the discrete wavelet transform spectrum estimator, the Fourier 
power spectrum  of the 1-dimensional (1-D) spatial distribution of these 
data is found to be almost flat on scales from 2 to about 100 h$^{-1}$ Mpc (
Pando \& Fang 1998). For comparison, Figure 1 also shows a randomized sample 
which is produced by a random shifting of the lines of QSO-0142. Obviously, 
one cannot simply distinguish the real sample and its random counterpart by 
inspection. In fact, when analyzed by second order statistics, the real data 
are still indistinguishable from randomized samples. 

\begin{figure}
\resizebox{\hsize}{!}{
\begin{turn}{-90}
\includegraphics{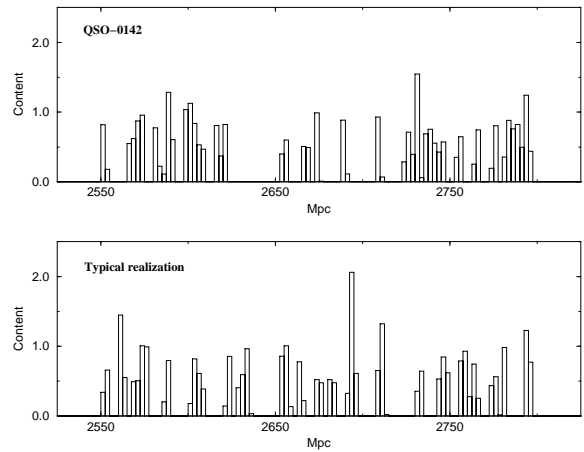}
\end{turn}
}
\caption[] { The Ly$\alpha$ forest line distribution of sample
   Q0142 and a typical realization of its randomization.}
\label{fig1}
\end{figure}

These results  indicate that 2nd order statistical techniques, i.e. the 
two-point correlation function and the power spectrum, are not
even qualitatively sufficient to describe the clustering features of these
samples. Higher order measures are not a correction to lower order
descriptions, but crucial in describing the Ly$\alpha$ forest traced 
matter field. 

This conclusion is strengthened by studying simulated samples.
Typically, simulated density fields for pre-collapsed clouds are
generated as perturbations with a linear or linear log-normal spectrum given
by models such as the cold dark matter model (SCDM), the cold plus hot
dark matter model (CHDM), and the low density flat cold dark matter model
(LCDM). The baryonic matter distribution is then produced by assuming that
the baryonic matter traces the dark matter distribution on scales larger than
the Jeans length of the baryonic gas, but is smooth over structures on scales
less than the Jeans length. The simulated Ly$\alpha$ absorption spectrum can
be calculated as the absorption of neutral hydrogen in the baryonic gas.
The effects of the observational instrumental point-spread-function,
along with Poisson and background noises can also be simulated
properly (Bi, Ge \& Fang 1995, hereafter BGF; Bi \& Davidson 1997).

Within a reasonable set of parameters, the simulated samples are found to
fit with observational measurements such as the number density of Ly$\alpha$
clouds, the distribution of equivalent widths, the red-shift-dependence of the 
width distributions and clustering. Regardless the details of the simulation, 
these samples should contain structures because the effects of gravitational
collapse have been considered, and baryonic matter does not distribute
uniformly random, but traces the structure of dark matter. However, as is the 
case with observations, the simulated samples show no power in their two-point
correlations (see Figure.11 of BGF), and their 1-D spectra are
rather flat on scales less than
100 h$^{-1}$Mpc. These results clearly show that higher order measures are 
necessary in describing the statistical features of the 1-D distributions of 
the Ly$\alpha$ clouds. 

\section{Higher Order Statistics via the Wavelet Coefficients}

\subsection{DWT decomposition and 2nd Order Statistics}

It is generally believed that the cosmic mass (or number) density 
distribution ${\rho(x)}$ can be mathematically treated as a homogeneous
random field. It is often more convenient to express ${\rho(x)}$ in terms
of its Fourier transform, $\rho(k)$, which is the Fourier coefficient of 
$\rho(x)$.

One reason for expressing the density distribution in terms of its Fourier
transform is that for Gaussian random fields all the statistical 
features of $\rho(x)$ can be completely described by the amplitude of the 
Fourier coefficients. In this case, the phase of $\rho(k)$ is not important and
the power spectrum of the perturbations, $|\rho(k)|^2$, or  
equivalently, the two-point correlation function, are all that is necessary
to describe the statistical 
behavior of the density distribution. However, if the field $\rho(x)$ 
is non-Gaussian, then in order to have a full statistical description of the
field the phases of the Fourier coefficients $\rho(k)$ are essential.

As is well known, it is difficult, even practically impossible, to find
information about the phases of the Fourier coefficient as soon as there is
some computational noise. The lack of information about the phases makes 
the $\rho(k)$ description incomplete: we might know the scales 
$k$ of the structures, but nothing about the positions of the considered 
structures. A possible way of simultaneously describing the scale and 
position of structures is provided by the discrete wavelet transform (DWT) 
(Daubechies 1992, Meyer 1993.)
The DWT analysis has successfully been applied to the problems of turbulence 
(Yamada \& Ohkitani 1991; Farge 1992) and high energy physics (Huang et al. 
1996.)  Our previous studies also show that the DWT analysis is 
useful for large scale structure study (Pando \& Fang 1996, Fang \&
Pando 1997, 1998; Pando et al. 1998, Fang, Deng \& Xia, 1998, Xu, Fang
\& Wu, 1998.) 

Let's consider a 1-D mass density contrast 
$\delta(x)=[\rho(x)-\bar{\rho}]/\bar{\rho}$, which covers a spatial range 
$0 \leq x \leq L$. The expansion of the field $\rho(x)$ in terms of the
DWT basis is given by
\begin{equation}
\delta(x) = \sum_{j=0}^{\infty} \sum_{l= 0}^{2^j -1}
  \tilde{\delta}_{j,l} \psi_{j,l}(x)
\end{equation}
where $\psi_{j,l}(x)$, $j=0,1,...$, $l=0...2^j-1$ are the basis of the DWT.
The DWT  basis are orthogonal and complete. The wavelet function
coefficient (WFC), $\tilde{\delta}_{j,l}$, is computed by
\begin{equation}
\tilde{\delta}_{j,l}= \int \delta(x) \psi_{j,l}(x)dx.
\end{equation}

The wavelet transform basis functions $\psi_{j,l}(x)$ are generated from the 
basic wavelet $\psi(x/L)$ by a dilation, $2^j$, and a translation 
$l$, i.e. 
\begin{equation}
\psi_{j,l}(x) =\left( \frac{2^j}{L} \right)^{1/2} \psi(2^jx/L-l).
\end{equation}
The basic wavelet $\psi$ is designed to be continuous, admissible 
and localized. Unlike the Fourier basis $\exp (- i2\pi nx/L)$, which are 
non-local in physical space, the wavelet basis $\psi_{j,l}(x)$ are 
localized in both physical space and  Fourier (scale) space. In physical 
space, $\psi_{j,l}(x)$ is centered at position $lL/2^j$, and in Fourier 
space, it is centered at wavenumber $2\pi \times 2^j/L$. Therefore,
the WFCs, $\tilde{\delta}_{j,l}$, have two subscripts $j$ and $l$, 
which describe, respectively, the scale and position of the density 
perturbations.

A clearer picture of how the transforms work can be seen in the phase space 
$(x,k)$. A complete, orthogonal basis set
resolves the whole phase space into ``elements" of size  
$\Delta x$ and $\Delta k$. Each mode corresponds to elements
in the  phase space. For the Fourier transform, this 
corresponds to elements of size  $\Delta k=0$ and 
$\Delta x = \infty$. For the wavelet transform, both 
$\Delta k$ and $\Delta x$ are finite, and the corresponding area of the
element is as small
as $\Delta x \Delta k = 2\pi$. That is, the DWT is able to resolve an arbitrary
function simultaneously in terms of $x$ and $k$ up to the limit of the 
uncertainty principle. The DWT decomposes the density fluctuation 
field $\delta(x)$ into domains ${j,l}$ in phase space, and for each mode,
the corresponding area in the phase space is as small as that allowed by the
uncertainty principle. 

The WFC $\tilde{\delta}_{j,l}$ and its intensity $|\tilde{\delta}_{j,l}|^2$ 
describe,
respectively, the fluctuation of the density and its power on scale $L/2^j$ at 
position $lL/2^j$. As with  the Fourier basis, Parseval's theorem 
holds for the DWT basis. It is (Fang \& Pando 1997, Pando \& Fang, 
1998)
\begin{equation}
\frac{1}{L} \int_{0}^{L} |\delta(x)|^2 dx = \sum_{j=0}^{\infty} 
\sum_{l = 0}^{2^j-1} |\tilde{\delta}_{j,l}|^2.
\end{equation}
It is possible to define the power spectrum of the density perturbation
on scale $j$ by the variance of the WFCs as
\begin{equation}
P_j =  \sum_{l = 0}^{2^j -1} 
(\tilde{\delta}_{j,l}-\overline{\tilde{\delta}_{j,l}})^2,
\end{equation}
where
\begin{equation}
\overline{\tilde{\delta}_{j,l}}\equiv \frac{1}{2^j}
 \sum_{l=0}^{2^j-1} \tilde{\delta}_{j,l}.
\end{equation}
It has been shown that the DWT power spectrum eq.(6) can be converted to the 
Fourier power spectrum, i.e. in terms of second order statistical description 
the DWT and Fourier transform are equivalent.

\subsection{One-point distribution of WFCs and non-Gaussianity}

The cosmic density field is usually assumed to be ergodic: the average over
an ensemble is equal to the spatial 
average taken over one realization. This is the so-called ``fair sample 
hypothesis" (Peebles 1980). A homogeneous Gaussian field with continuous 
spectrum is certainly ergodic (Adler 1981). In some non-Gaussian cases, 
such as homogeneous and isotropic turbulence (Vanmarke, 1983), ergodicity 
also approximately holds. Roughly, the ergodic hypothesis is reasonable if 
spatial correlations are decreasing sufficiently rapidly with increasing 
separation. In this case, the volumes separated by large distances are 
approximately statistically independent, and can be treated as independent 
realizations. Note that the $\psi_{j,l}(x)$ are orthogonal with 
respect to the position index $l$, and therefore, for an ergodic field, 
the $2^j$ WFCs, $\tilde{\delta}_{j,l}$, $l=0,1...2^j-1$, at a given $j$ 
should be statistically independent. Thus the WFCs 
$\tilde{\delta}_{j,l}$ at fixed $j$ and different $l$ can be treated as
independent measures of the density fluctuation field. The $2^j$ WFCs, 
$\tilde{\delta}_{j,l}$, 
from {\it one} realization of $\delta(x)$ can be employed as a statistical
ensemble. In this way, when the fair sample 
hypothesis holds, an average of the ensemble can be well estimated 
by averaging over $l$, i.e. $\langle \tilde{\delta}_{j,l} \rangle \simeq 
(1/2^j)\sum_{l=0}^{2^j-1}\tilde{\delta}_{j,l}$,
where $\langle ...\rangle$ denotes the ensemble average.

Consequently, the distribution of the $\tilde{\delta}_{j,l}$   
is actually the one-point distribution of the WFCs at a given 
scale $j$. The non-Gaussianity of the density field $\delta(x)$ can directly 
be measured by the deviation of the one-point distribution from a Gaussian 
distribution. For this purpose, we can calculate the cumulant moments 
defined by (Carruthers 1991; Carruthers, Eggers \& Sarcevic 1991)
\begin{equation}
I^2_j= M^2_j,
\end{equation}
\begin{equation}
I^3_j = M^3_j,
\end{equation}
\begin{equation}
I^4_j= M^4_j - 3 M^2_j M^2_j,
\end{equation}
\begin{equation}
I^5_j= M^5_j - 10 M^3_jM^2_j,
\end{equation}
where
\begin{equation}
M^n_j \equiv \frac{1}{2^j} \sum_{l=0}^{2^j-1}
(\tilde{\delta}_{j,l} - \overline{\tilde{\delta}_{j,l}})^n.
\end{equation}
From eqs.(6), (7) and (8), one sees that the second order cumulant 
moment is the DWT power spectrum on the scale $j$, i.e. 
\begin{equation}
I^2_j = \frac{1}{2^j}P_j.
\end{equation}

For Gaussian fields all the cumulant moments higher than order 2
are  zero. Thus one can measure the non-Gaussianity of
the density field $\delta(x)$  by $I^n_j$ with $n>2$.
Analogous to $I^2_j$ being called the DWT power spectrum, we will
call $I^n_j$ the DWT spectrum of $n$-th cumulant. The cumulant measures
$I^3_j$ and $I^4_j$ are related to the better known skewness and
kurtosis, respectively. Thus, the non-Gaussianity of $\delta(x)$ can be 
detected by the DWT skewness and kurtosis spectra defined as
\begin{equation}
S_j \equiv \frac{1}{(I_j^2)^{3/2}} I^3_j,
\end{equation}
\begin{equation}
K_j  \equiv   \frac{1}{(I_j^2)^2} I^4_j.
\end{equation}
$S_j$ and $K_j$ are basic statistical measures employed in this paper.

For comparison, the definitions of the ``standard''  skewness $S$ and 
kurtosis $K$ for a 1-D distribution $\delta(x)$ covering on
$N$ bins are given as follows
\begin{equation}
S \equiv \frac{1}{N \sigma^3} \sum_{i=1}^{N}
(\delta(x_i) - \overline{\delta(x_i)})^3,
\end{equation}
and
\begin{equation}
K \equiv \frac{1}{N\sigma^4} \sum_{i=1}^{N}
[(\delta(x_i) - \overline{\delta(x_i)})^4] - 3,
\end{equation}
where the variance $\sigma^2$ is given by
\begin{equation}
\sigma^2 \equiv \frac{1}{N -1} \sum_{i=1}^{N}
(\delta(x_i) - \overline{\delta(x_i)})^2.
\end{equation}
Obviously, no scale information is given by $S$ and $K$.

\subsection{Central limit theorem and the DWT analysis}

It is well known that not all one-point distributions can detect 
non-Gaussianities . This is due to the constraints imposed by the
central limit theorem. For instance, if the universe consists of a large 
number of dense clumps with a non-Gaussian probability distribution function 
(PDF), the one-point distributions of the real and imaginary components of each
individual Fourier mode are still Gaussian due to the central limit theorem
(Ivanonv \& Leonenko 1989). Further, even when the non-Gaussian clumps are
correlated the central limit theorem still holds if the two-point correlation
function of the clumps approaches zero sufficiently fast (Fan \& Bardeen
1995). For these reasons, the one-point distribution function of Fourier
modes is not sensitive enough to detect deviations from Gaussian
behavior. Even for samples with strong non-linear evolution, the one point
distribution function of  Fourier modes is found to be consistent with
a Gaussian distribution (Suginohara \& Suto 1991). It should be pointed out
that  the inefficiency of the Fourier mode one-point distribution in
detecting non-Gaussianity is not because Fourier transform loses information
about the distribution $f(x)$. The Fourier coefficients contain all the
information on non-Gaussianity, but the information is mainly contained in
the phases of the Fourier coefficients. As mentioned in \S 3.1, it is very 
difficult to detect the distribution of the phases of Fourier coefficients.

On the other hand, the wavelet coefficients are not subject to the central
limit theorem. In this respect, the DWT analysis is similar to the count in
cell (CIC) method (Hamilton 1985; Alimi, Blanchard \& Schaeffer 1990;
Gazta\~naga \& Yokoyama 1993; Bouchet et al. 1993; Kofman et al. 1994;
Gazta\~naga \& Frieman 1994). The CIC is not subject to the central limit
theorem as it based on spatially localized window functions (Adler 1981). The
DWT's basis are also localized. If the scale of a clump in the universe is
$d$, eq.(3) shows that the WFCs, $\tilde{\delta}_{j,l}$, with $j=\log_2(L/d)$,
is determined only by the density field in a range containing no more than
one clump. That is, for scale $j$ the wavelet coefficients are not given by
a superposition of a large number of clumps, but are determined by at most
one of them. Therefore, it avoids the constraints imposed by the central
limit theorem.

This point can also be shown from the orthonormal basis being used for the
expansion of the distribution $\delta(x)$. A basic condition of the central
limit theorem is that the modulus of the basis be less than $C/\sqrt L$,
where $C$ is a constant (Ivanonv \& Leonenko 1989). Obviously, all
Fourier-related orthonormal basis satisfy this condition because 
$(1/\sqrt L) |\sin kx| < C/\sqrt L$ and
$(1/\sqrt L)|\cos kx| < C/\sqrt L$, where $C$ is independent of coordinates in
both physical space $x$ and scale space $k$. On the other hand, the
normalized wavelets functions have [eq.(4)]
\begin{equation}
|\psi_{j,l}(x)| \sim \left( \frac{ 2^{j}}{L}\right )^{1/2} {\rm O}(1).
\end{equation}
Because the magnitude of the basic wavelet $\psi(x)$ is of order 1. The
condition  $ |\psi_{j,l}(x)|< C/ \sqrt{L}$ will no longer hold for a constant
$C$ independent of scale variable $j$. Hence, the one-point-distribution of
wavelet coefficients on scale $j$ should be a good estimator for the PDF of
a distribution. 

\section{Skewness and kurtosis spectra of Ly$\alpha$ forests}

\subsection{Data preparation}

The observational QSO Ly$\alpha$ forests data have a complex
geometry. By this we mean that different forests cover different spatial 
ranges, and no one of the forests distributes on the entire range  
$(D_{min}, D_{max})$. At the very least, a complicated weighting scheme is 
needed in order to form an ensemble. The DWT provides a way around this  
problem. Because wavelets are localized, the wavelet coefficients are only 
determined by 
the local distribution. If a forest sample lies in range $(D_1, D_2)$, one 
can extend it to $(D_{min},D_{max})$ by adding zeros to the data in ranges 
$(D_{min}, D_1)$ and $(D_2,D_{max})$. The wavelet coefficients in the
interval  $(D_1, D_2)$ will not be affected by the addition of zero in
$(D_{min},D_1)$ and $(D_2, D_{max})$. Any statistics can then be computed by 
the ensemble of wavelet coefficients, $\tilde\delta_{j,l}$, by simply
dropping  all wavelet coefficients with $l$ in the added zero ranges.

More precisely, the effect of zero padding on the wavelet coefficients is 
described by the so-called ``influence cone" which consists of the
spatial support of all dilated wavelet functions. For instance, if 
$\psi_{jl}(x)$ is localized in the space interval $\Delta x$ for 
$j=0$, the influence cone centered at $x_0$ is given by 
$x \in [x_0 - (\Delta x/2^{j+1}), x_0 + (\Delta x/2^{j+1})]$ (Farge 1992). 
Wavelet coefficients corresponding to positions outside
$(\Delta x/2^{j+1})$ will not corrupt information within the
influence cone.

The LWT and JB samples cover a red-shift range of 1.7 to 4.1, i.e. the 
spatial range in comoving distance is from about $D_{min}$=2,300 $ 
h^{-1}$Mpc to $D_{max}=$3,300 $h^{-1}$Mpc, if $q_{0} = 1/2$ and 
h$=H_0/100$ km s$^{-1}$ Mpc$^{-1}$.
To eliminate the proximity effect, all lines with $z \geq z_{em} - 0.15$ are 
deleted from our samples.  These two samples are extended to 
$(D_{min}, D_{max})$ as described above,
and binned into in 512 bins with comoving size $\sim$ 2.5 h$^{-1}$ Mpc, which 
is about the scale where the effect of line blending occurs. If a line is 
not located at the center of a bin, it is separated into the two neighboring 
bins, weighted according to the distance to each of the centers. 

The HKCSR and KT samples (HKCSR+KT),  can be treated the same 
way.  Since HKCSR and KT cover a smaller red-shift range from 
about 2.4 to 3.1, the forests of HKCSR+KT are extended to 128 bins  
of comoving size $\sim$ 2.5 h$^{-1}$ Mpc. 
Thus it is possible to compare all 3 sets for scales down to $j = 6$.
 
\subsection{Skewness and kurtosis spectra of real data}

As opposed to  $\delta(x)$ discussed in \S 3,  real data provide only
histograms of Ly$\alpha$ line distributions, i.e. the data are not a 
continuous function of $x$, but a point process in $x$ space. In this case, 
the WFCs $\tilde{\delta}_{j,l}$ will not be calculated by an integral like
eq.(3), but by the wavelet transformation matrix (Press et al. 1992.)

\begin{figure}
\resizebox{\hsize}{!}{
\begin{turn}{-90}
\includegraphics{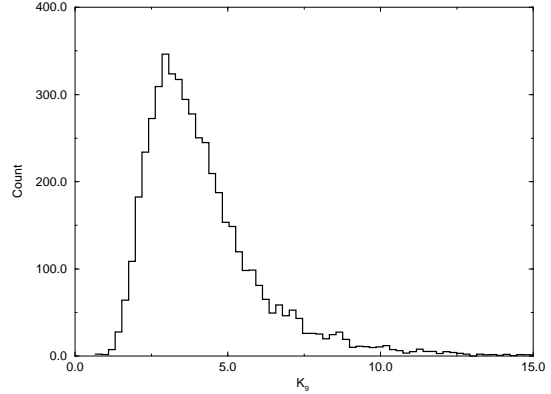}
\end{turn} }
\caption[]{Frequency distributions of $K_9$ from 5000 random
samples. It can be clearly seen that this distribution is non-Gaussian.
}
\label{fig2}
\end{figure}

It is well known that point processes generally are non-Gaussian.
For instance, if the Ly$\alpha$ clouds do not
distribute structurally, then $\delta(x)$ should be a 1-D Poisson process.
However, Poisson processes are non-Gaussian. Obviously, this non-Gaussianity
is not what we are seeking to measure because it is not the result of
non-linear evolution or physical processes related to structure formation.
Therefore, we should carefully distinguish the clustering related
non-Gaussianity with that from sampling and binning. For this purpose, we 
generate 5000 random realizations in which the mean number 
density follows eq.(1). Each realization is  
treated in the same way as the real data sets. From these random 
samples we can produce frequency distributions for each statistic, say 
$K_9$ (Figure 3). These frequency distributions are fair estimators of the 
underlying probability distribution. It can clearly be seen from Figure 2 that 
the random data are non-Gaussian. 

\begin{figure}
\resizebox{\hsize}{!}{
\begin{turn}{-90}
\includegraphics{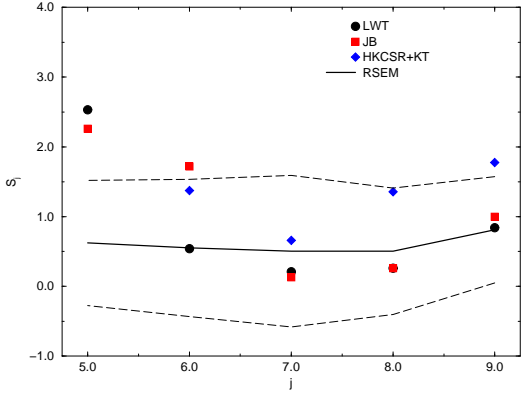}
\end{turn}
}
\caption[] {Skewness spectrum for LWT ($W>0.36 $\AA),
JB ($W>0.32 $\AA) and  HKCSR+KT samples. The red-shift evolution model (RSEM)
given by eq. (1), and the 95\% confidence limits are also shown. The
physical scale is related to $j$ by $2.5\times 2^{10-j}$ h$^{-1}$ Mpc.}
\label{fig3}
\end{figure}

Using these frequency distributions, the confidence levels for the mean values
of the random data can be estimated.  These are then to be compared with the
real data.

\begin{figure}
\resizebox{\hsize}{!}{
\begin{turn}{-90}
\includegraphics{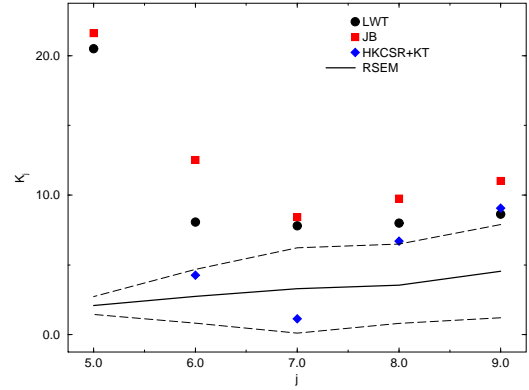}
\end{turn} }
\caption[]{Kurtosis spectrum for LWT ($W>0.36 $\AA),
JB ($W>0.32 $\AA) and  HKCSR+KT samples. The red-shift evolution model (RSEM)
given by eq. (1), and the 95\% confidence limits are also shown.}
\label{fig4}
\end{figure}

\begin{figure}
\resizebox{\hsize}{!}{
\begin{turn}{-90}
\includegraphics{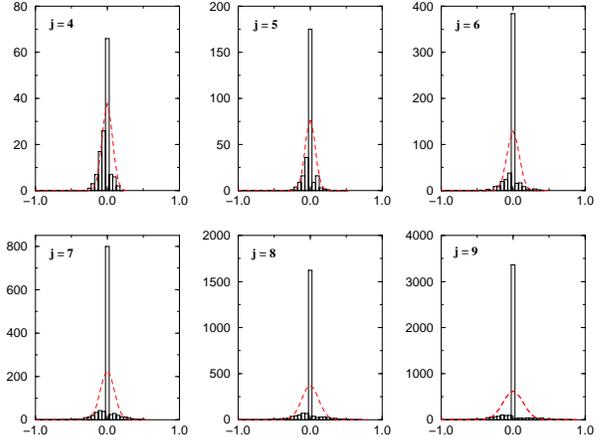}
\end{turn}
}
\caption[b]{ Histogram of one-point distribution of wavelet
coefficients for JB (W$>0.32$\AA).  At each scale $j$, the Gaussian
distribution (dashed line) has the same variance and normalization as the
wavelet coefficient distribution.}
\label{fig5}
\end{figure}

The skewness and kurtosis spectra of the LWT (W$>0.36$ \AA), JB
(W$>0.32$ \AA) and HKCST+KT (b $> 78$ km s$^{-1}$) data are shown in 
Figures 3 and 4, respectively. The skewness and kurtosis spectra of random
data and the  95\% confidence levels are also shown in these figures.
Figures 3 and 4 show that the skewness and the kurtosis spectra of the LWT
(W$>0.36 $\AA) and JB (W$>0.32 $\AA) are almost the same. The kurtosis
spectra are $>95\%$, significantly different from the random samples 
on all scales, and, on scales $j$= 5, the skewness spectra are also 
$>95\%$ and significantly different from the random samples.  
The Keck data, i.e.  HKCSR+KT, have qualitatively the same non-Gaussian 
behavior, 
especially at $K_8$ and $K_9$, where the HKCSR+KT data are the
same as that of LWT and JB. The HKCSR+KT gives lower values for $K_6$
and $K_7$ than the LWT and JB, while it gives higher $S_8$ and $S_9$
than LWT and JB. With the current data it is not possible to determine whether 
these differences are due to the higher resolution of the HKCSR+KT
data. It is important to realize that these observational data sets
were compiled by different groups, on different instruments, and as much as
6 years apart. These data sets are very independent and makes more convincing
the case for the existence of the non-Gaussianity at least on scales of
5 - 10 h$^{-1}$ Mpc, and confirms that the features shown in the
non-Gaussian spectrum are intrinsic features of the density field traced by
Ly$\alpha$ forests.

We can directly describe the non-Gaussianity by the one-point distributions
of the WFCs. Figure 5 plots the one-point distributions of
$\tilde{\delta}_{j,l}$ for the JB sample with $>0.32 $\AA. For each scale
$j$, the corresponding Gaussian distribution is plotted such that it has the
same normalization and variance as the one-point distribution. This figure
clearly shows that all the distributions on scales $j>4$ (or less than about
80 h$^{-1}$ Mpc) are significantly non-Gaussian. These distributions are also
asymmetric, with fewer positive wavelet coefficients.

This asymmetry is at least partially due to the red-shift-dependence 
of the Ly$\alpha$ clouds. The wavelet coefficient $\tilde{\delta}_{j-1,l}$ is
mainly determined by the difference of (positive) densities at $\{j, 2l\}$ 
and $\{j, 2l+1\}$ (PF). For a clump in red-shift space, the density 
change on the lower red-shift or lower $l$ side contributes negative wavelet 
coefficients, while the higher red-shift side gives positive wavelet
coefficients. If as shown in PF, the number of Ly$\alpha$ clumps decreases
with increasing red-shift, the change in clustering amplitudes (wavelet
coefficients) on the higher red-shift side (positive wavelet coefficients)
should be less than the lower side (negative wavelet coefficients). That is,
the number of positive wavelet coefficients will be less than negative
wavelet coefficients. This is consistent with a small, but positive skewness.

\subsection{Removing degeneracy by the non-Gaussian spectrum}

We have shown that cluster identification by a wavelet decomposition
is a useful tool for discriminating among models of Ly$\alpha$ clouds
(PF). The spectrum of non-Gaussianity can play the same role. Namely, 
$S_j$ and $K_j$ are effective measures for removing the degeneracy that
exists at second order among models.

As an example, we examined the BGF simulated Ly$\alpha$ forest samples.
This simulation shows that  two 
models, SCDM and LCDM, are degenerate if only the first (number
density) and second (variance, or power spectrum) order statistics are 
considered. That is, both SCDM and LCDM give about the same predictions for
the following features of the  Ly$\alpha$ forests: a.) the number density of
Ly$\alpha$ lines and its dependencies on red-shift and equivalent width; b.)
the distribution of equivalent widths and its red-shift dependence; c.) the
two-point correlation function. 

\begin{figure}
\resizebox{\hsize}{!}{
\begin{turn}{-90}
\includegraphics{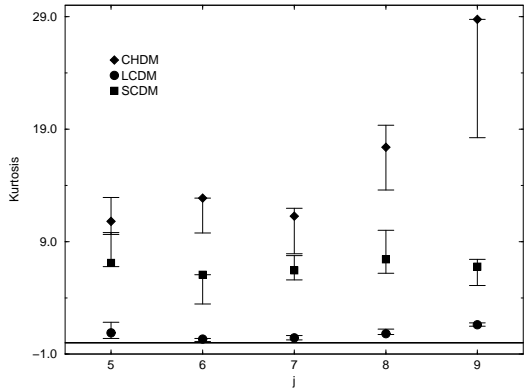}
\end{turn}
}
\caption[]{Kurtosis spectrum of BGF samples in the SCDM, LCDM,
and CHDM models with $W \ge 0.32$\AA.}
\label{fig6}
\end{figure}

This degeneracy can be removed by the non-Gaussian spectrum. Figure 6 plots
the kurtosis spectra for the LCDM, SCDM, and CHDM models, in which the 
Ly$\alpha$ lines are chosen with width
$W>0.32$\AA. The error bars in Figure 6 are given by the distribution of
20 realizations for each model. 

Figure 6 shows that even though the BGF simulation is based on the linear
power spectrum of the density perturbations, the Ly$\alpha$ distribution is
non-Gaussian. This is because the selection of peaks from the density field
is a non-Gaussian processes. More important, Figure 7 clearly shows the
significant difference among the three models. With 95\% confidence, the 
$K_j$ amplitudes of all three models are different
on all scales. The $K_j$ of the SCDM model is significantly larger than
zero for all scales $j$. Yet, the LCDM model gives much lower $K_j$ for all 
scales
$j$. The non-Gaussian spectrum provides a extremely effective method
for removing the degeneracy among models.

\section{Discussion and conclusions}

The DWT skewness and kurtosis spectra of real and simulated samples of
Ly$\alpha$ forests has been calculated. Deviations from a Gaussian state 
in these samples are detected for all data on scales of 5 to 10 h$^{-1}$ 
Mpc with confidence level larger than 95\%. For the two sets which are 
available on large scales, the non-Gaussianity is significant even on scales 
as large as about 80 h$^{-1}$ Mpc. The scale-dependence (or spectrum) for 
a The higher order DWT cumulant is especially effective in distinguishing the 
real samples from random data, and in  discriminating among models. 

It is possible for non-Gaussianity to result from systematic 
errors hiding in the original data reduction. For instance, discrete sampling 
or binning of a continuous distribution into a histogram with bin size 
$\Delta D$ leads to non-Gaussianity at least on scale $\Delta D$. Many selection 
effects actually are some kind of sampling, and therefore, they will also give 
rise to non-Gaussianity. All these non-Gaussianities are ``spurious".

Fortunately, in most cases these spurious non-Gaussian signals are
significant only at one special scale. For a Poisson process, it is the scale 
of the mean distance of nearest neighbor clouds. For binning, it is $\Delta D$.
On scales larger than $\Delta D$, the spurious non-Gaussianities will 
decay out. That is, the scale-dependent behavior of sampling and binning 
is useful in recognizing spurious non-Gaussianity.
In the case of the Ly$\alpha$ forests, the mean nearest neighbor distance
along with the binning scale $\Delta D$ are much less than the scales of the
detected non-Gaussianities, and therefore, the detected non-Gaussianities are 
inherent to the distributions.  There is also evidence for intrinsic
scales in the Ly$\alpha$ spectra of magnitude $\sim$ 25 h$^{-1}$ Mpc 
(Rauch 1998).  However this contamination can be reduced by studying
$K_j$ and $S_j$ on scales different from 25 h$^{-1}$ Mpc.
 
Nevertheless, high-resolution samples which can cover the spatial range as l
arge as 80 h$^{-1}$ Mpc are needed before a final determination can be made.  
Recently a scale-scale correlation analysis also shows a non-Gaussianity in the 
Ly$\alpha$ line distribution on such large scales (Pando et al. 1998). It seems to 
support our conclusion that cosmic matter underwent a non-Gaussian process on 
quite large scales at quite early times.

Last but not least, the numerical work of calculating DWT higher order
moment spectrum is not any more difficult than calculating
the second order moments. Generally, the calculation of third and higher
order correlation functions of large scale structure samples is very 
strenuous work.
But the numerical work involved in calculating the DWT is about the same
as the FFT, and can even be faster. The FFT requires $\sim$ N Log N 
calculations, while the DWT, using a ``pyramid" scheme, requires only order
N calculations  (Press et al. 1992).

This method can easily be generalized to 2- and 3-dimensions.
The kurtosis and skewness spectrum opens a new window for looking at
the statistical features of large scale structures. It is an important and
necessary addition to the existing methods of describing the clustering and
correlation of the cosmic density field, and for discriminating among models
of structure formation.

\begin{acknowledgements}
Both authors wish to acknowledge the debt owed to the late Professor 
P. Carruthers who initiated the DWT study in our group.  He will be sorely
missed by us both. We also thank to Drs. H.G. Bi and P.Lipa for
insightful conversations.
\end{acknowledgements}

\end{document}